\documentclass[amsmath,amssymb,aps,reprint,superscriptaddress,longbibliography]{revtex4-1}
\usepackage{graphicx}
\usepackage{dcolumn}
\usepackage{bm}
\usepackage[titletoc,title]{appendix}
\usepackage[utf8]{inputenc}

\newcommand\p{\partial}

\newcommand\z{\zeta}

\newcommand\ta{\theta}

\newcommand\ph{\varphi}

\newcommand\eps{\epsilon}

\newcommand\di{\textrm{d}}
\newcommand\ex{\textrm{e}}
\newcommand\br{\textbf{r}}

\begin{document}

\preprint{APS/123-QED}

\title{Phase field model for phagocytosis dynamics  }

\author{Mohammad Abu Hamed}
\affiliation{Department of Mathematics, Technion - Israel Institute of Technology, Haifa 32000, Israel }
\affiliation{Department of Mathematics, The College of Sakhnin - Academic College for Teacher Education, Sakhnin 30810, Israel }

\author{Alexander A. Nepomnyashchy}
\affiliation{Department of Mathematics, Technion - Israel Institute of Technology, Haifa 32000, Israel }

\begin{abstract}
The basic process of the innate immune system when phagocyte (white blood cell) engulf or swallow a target particle
(bacterium or dead cell), is called phagocytosis.
We apply the phase field approach in the spirit of \cite{Winkler+Aranson+Ziebert2019}, that couples the order
parameter $u$ with 3D polarization (orientation) vector field $\textbf{P}$ of the actin network of the phagocyte
cytoskeleton. We derive a single closed scalar integro-differential equation governing the 3D phagocyte membrane
dynamics
during bead engulfment, which includes the normal velocity of the membrane, curvature, volume relaxation rate, a function determined by the
molecular effects of the subcell level, and the adhesion effect of the motionless rigid spherical bead. This equation
is easily solved numerically. The simulation manifests the pedestal and the cup phases but not the final complete bead
internalization.
\end{abstract}

\maketitle

\section{Introduction}
The process when a phagocyte, such as white blood cell, swallows or engulfs a target particle, such as dead cell or
bacterium, larger than $0.5\mathrm{\mu m}$ in diameters, is called phagocytosis. It plays a prominent role
in the innate immune system of an organism \cite{Flannagan+Jaumouille+Grinstein2012}.

The phagocyte has receptors at its membrane, while the target particle's membrane expose a molecular pattern. When
the
phagocyte is in the vicinity of a target particle, the phagocyte receptors bend to the molecules at the particle
surface. Various of chemical interactions between them will occurs. These interaction trigger "eat me" signal for the
dead cell or foreign cell, or "do not eat me" signal for the healthy cell \cite{Poon+Hulett+Parish2010}. In the first
case, a bio-chemical pathway  inside the phagocyte will generate in order to stimulate F-actin polarization in the
cytoskeleton, to form membrane protrusion, that later extends to pseudopodia that shapes the phagocytic cup phase. The
last step is the rounding phase, when the pseudopodias embrace the target particle, their edges fuse together to form the internalized phagosome \cite{Querol+Rosales2020}, \cite{Rosales+Querol2017}.

Several computational models describing the dynamics of antibody coated bead engulfed by professional phagocyte, where suggested in the literature
\cite{Herant+Heinrich+Dembo2006}, \cite{Richards+Endres2014}, \cite{Tollis+Dart+Tzircotis+Endres2010}, \cite{Zon+Tzircotis+Caron+Howard2009}. The remarkable work of \cite{Herant+Heinrich+Dembo2006}, \cite{Herant+Lee+Dembo+Heinrich2011} is the most realistic.
This is a deterministic model that considers the phagocyte as a mixture of two materials (phases), the cytoskeleton
and the cytosol, enclosed by the phagocyte membrane. The dynamics of this continuum model is governed by mass and momentum
conservation. This model is solved numerically via the finite element method. The model include adjustable parameters
iteratively optimized with experimental data. That model manifests the three basic sequential phagocytosis phases,
pedestal, cup, and the rounding phase, i.e., the complete bead internalization.

Another model was developed in \cite{Richards+Endres2014}. The authors consider the phagocyte membrane as an infinite
line with
distributed receptors and signals density located in the vicinity of a circular bead with distributed given
ligands. The model
governs the dynamics of the curve line around the bead and the one-dimensional density functions. It
neglects the details of the cup shape. By fitting the model parameters with experimental data, one observes that
phagocytosis occurs in two stages where the contact area obey different power laws. In addition, the
engulfment of bead shapes other than sphere is considered \cite{Richards+Endres2016}. Also, a three-dimensional
stochastic
biophysical
model of phagocytosis utilizing the zipper mechanism is developed in \cite{Tollis+Dart+Tzircotis+Endres2010}.

In the present paper we consider the phase field approach to model motionless rigid bead engulfment by phagocyte, in
the spirit of the model developed in \cite{Winkler+Aranson+Ziebert2019}, that describes 3D cell crawling on various
substrates topography. This is reasonable due to the similarity between the actin-based protrusion that developed
during phagocytosis and lamellipodium formation \cite{Jaumouille+Waterman2020}.

We design the initial configuration of phagocytosis as phagocyte sphere that is tangent to a fixed bead sphere.
Phagocytosis is the
interface dynamic of the initial phagocyte sphere while it surrounds the given rigid  bead sphere, see Fig \ref{schematic}. We consider a minimal phase field model that is a simplified version of that developed in \cite{Winkler+Aranson+Ziebert2019}, which couples the order parameter with 3D polarization (orientation) vector field of the actin network. Assuming cylindrical symmetry we derive a single closed integro-differential equation governing the phagocyte dynamics, that could easily solved numerically.
Relative to the previous mentioned models, our model is simple since we try to describe the complex
reality of phagocytosis by a close single scalar equation for the phagocyte membrane dynamics.

In the next section we present the minimal 3D phase field model. We
introduce the proper length and time scales for the phagocytosis dynamics. We perform asymptotic analysis and find the
fields at the leading order, and then we use the solvability condition to derive a closed evolutionary nonlocal
equation that describes the phagocyte interface dynamics \eqref{eveq}. This equation is solved numerically via the function
\verb"NDSolve" of Wolfram Mathematica. Finally, we present the conclusions.

%
%
%
%

\section{Formulation of the problem}
\begin{figure}
  \centering
  \includegraphics[scale=0.3]{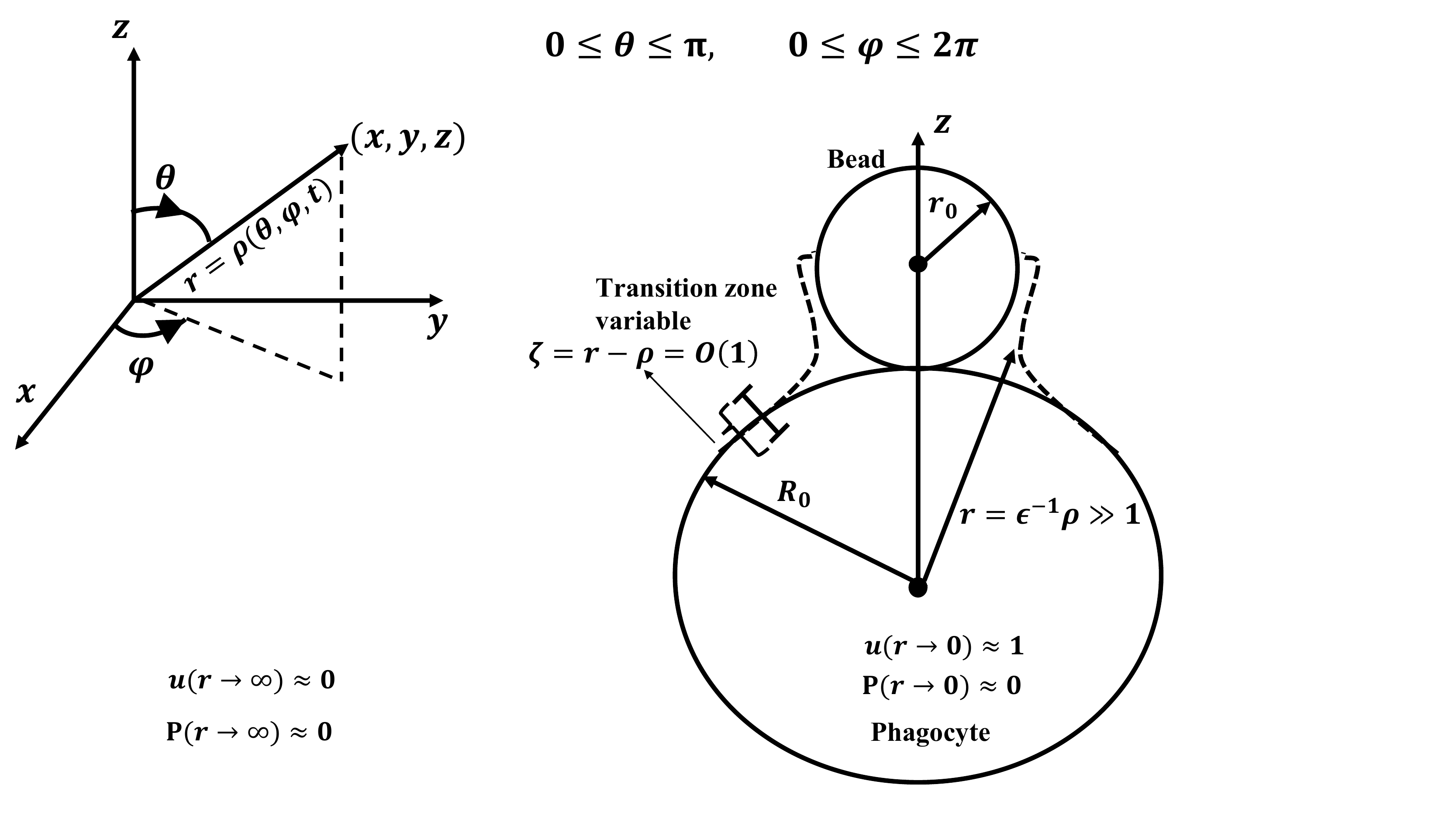}
  \caption{A schematic description of fixed rigid spherical bead engulfment by phagocyte. We present the boundary conditions of the model \eqref{Mod}, and the transition zone variable. Also we present the description of the standard
spherical coordinates.    } \label{schematic}
\end{figure}

For understanding phagocytosis we consider the scenario of an initial phagocyte sphere of radius $R_0$  tangent to a
motionless rigid spherical bead of radius $r_0=\lambda R_0$. The phagocytosis phenomenon is modeled by the dynamics
of a phagocyte surface engulfing the bead sphere, see Fig. \ref{schematic}.

Since phagocytosis is an actin based cell motility \cite{May+Machesky2001}, we consider the following simplified
version of the minimal model that was developed in \cite{Winkler+Aranson+Ziebert2019}.
Each equation and term in this model is explained and discussed in details by the authors in \cite{AbuHamed2021}.
\begin{subequations}\label{Mod}
\begin{eqnarray}
&& u_t = D_u \nabla^2 u  -(1-u)(\delta-u)u - \nonumber \\
&& \alpha \nabla u \cdot \textbf{P} - k \nabla \Psi_u \cdot \nabla u ,\label{Mod1} \\
&& \delta = \frac{1}{2} + \mu \delta V -\sigma |\textbf{P}|^2,\ \delta V(t)= \int u \di^3 r - v_0, \label{Mod2}\\
&& \textbf{P}_t = D_p \nabla^2 \textbf{P}  - \tau^{-1}\textbf{P}  - \nonumber \\
&& \beta\Psi_p(\br) \left[(1-\nu)\hat{P} \nabla u + \nu\nabla u \right] ,\label{Mod3}\\
&& \quad u(r\rightarrow\infty)=0,\label{Mod4}\\
&& \textbf{P}(r\rightarrow\infty)=0;\label{Mod5}
\end{eqnarray}
\end{subequations}
 Here $u$ is the order parameter that is close to $1$ inside the cell and $0$ outside.
The three-dimensional polarization vector field $\textbf{P}$ representing the actin orientations is generated by the
inhomogeneity of $u$, therefore it is small everywhere except the cell boundary;
$\hat{P}= \hat{I}-\hat{n}\hat{n}$ is the projection operator onto the local tangential plane of the spherical bead
substrate, $\hat{n}$ is the normal vector to the substrate, therefore $ \hat{P} \hat{n} =0$.

The constant parameters of the problem are: $D_u$ is the stiffness of diffuse interface, $D_p$ is the diffusion
coefficient for $\textbf{P}$, $\alpha$ is the coefficient characterizing advection of $u$ by $\textbf{P}$, $\beta$ determines the creation of $\textbf{P}$ at the interface, $\tau^{-1}$ is the inverse time of the degradation of $\textbf{P}$ inside the cell, $v_0$ is the overall initial volume of the cell, $\mu$ is
the stiffness of the volume constraint, and $\sigma$ is the contractility of actin filament bundles. All the parameters listed above are positive.

In addition, we make the basic assumption that the ratio $\eps$ of the thickness of the cell wall (i.e., the width of
the transition zone, where $u$ is changed from nearly 1 to nearly 0) to the characteristic size of the cell is small,
$\eps\ll 1$, see Fig. \ref{schematic}.

In order to describe the engulfment around a motionless spherical rigid bead, motivated by
\cite{Winkler+Aranson+Ziebert2019}, we define the static fields as
\begin{subequations}\label{Psi}
\begin{eqnarray}
 && \Psi_u (\textbf{r} ) = \exp  \left[ -\eps^2 \Big( |\textbf{r} - (r_0+R_0)\hat{z}| -r_0 \Big)^2 /D_u \right] , \\
 && \Psi_p (\textbf{r})= \exp  \left[ -\eps^2 \Big( |\textbf{r} - (r_0+R_0)\hat{z}| -r_0 \Big)^2 /(\tau D_p) \right],
\end{eqnarray}
\end{subequations}

\begin{figure}
  \centering
  \includegraphics[scale=0.3]{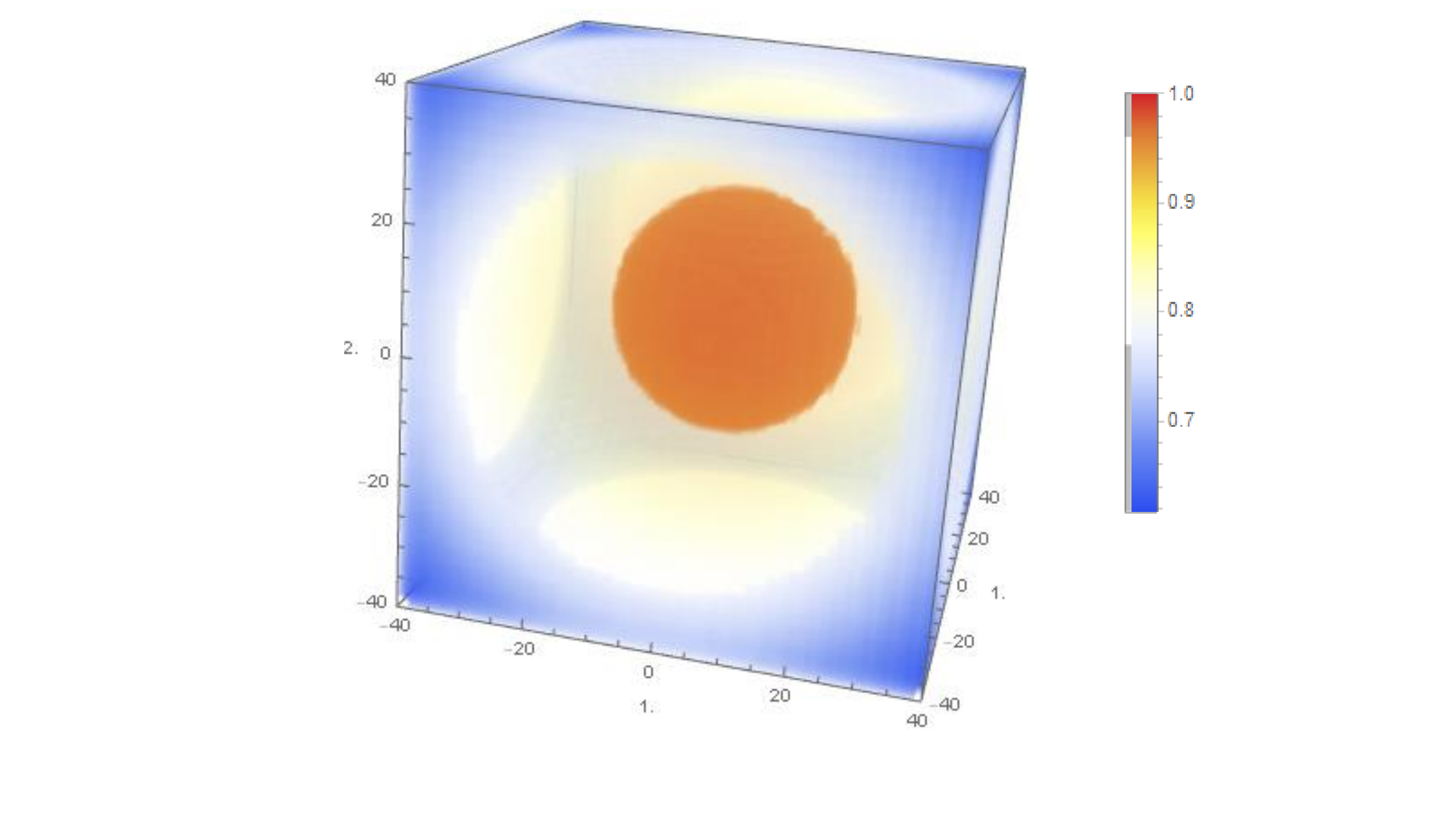}
  \caption{ Plot of the density functions $\Psi_{u}(\textbf{r})$ in \eqref{Psi}. Notice that $\Psi_{u}=O(1)$ nearby the spherical bead  (red region) while attenuate far away both inside and outside   (blue region).The width of the layer where $\Psi_u=O(1)$ is $O(\sqrt{ D_u})$ .   The density plot for $\Psi_p(\textbf{r})$ is similar.  } \label{density}
\end{figure}

see Fig. \ref{density}. We choose $\tau D_p>D_u$ to allow more substantial actin inside the cell. The appearance of $\eps$
in the exponents of \eqref{Psi} allows to avoid boundary layer problem complications both in time and space. As we
see, the functions $\Psi_u$ and $\Psi_p$ are $O(1)$ in the vicinity of the spherical bead and exponentially decay
otherwise.

The expression $k \nabla \Psi_u \cdot \nabla u$ in \eqref{Mod1} models the adhesion effect of the bead substrate; $k$
is  adhesion strength parameter. Notice that $\nabla \Psi_u \cdot \nabla u =O(1)$ only in region near the bead and also at the cell boundary or membrane i.e., where protrusion  holds.
  Also, the appearance of $\Psi_p \nabla u$  in \eqref{Mod3} allows high actin concentration near the bead substrate
where protrusions and pseudopodia are developed during phagocyte membrane morphology remodeling, forming the cup phase,  and low actin concentration otherwise, see Fig. \ref{schematic}.

We apply the standard spherical coordinate system,
\begin{subequations}\label{coor1}
\begin{eqnarray}
&&(r,\ta,\ph), \ 0<r<\infty, \ 0<\ta< \pi, \ 0<\ph<2\pi,\\
&& x=r\sin\ta \cos\ph, \ y= r\sin\ta \sin\ph, \ z=r\cos\ta,
\end{eqnarray}
\end{subequations}
 see Fig. \ref{schematic}, and assume a cylindrical symmetry of
the problem, i.e., independence of all the fields on the azimuthal angle $\ph$. Hence $u=u(r,\ta)$,
$\textbf{P}(r,\ta,t)=p\hat{r}+q\hat{\ta}$, and
\begin{subequations}\label{coor2}
\begin{eqnarray}
&& \nabla = \hat{r}\p_r + \hat{\ta}\frac{\p_\ta}{r} ,\\
&& \nabla^2 = \p_{r}^2 + \frac{2\p_r}{r}+ \frac{1}{r^2}\Big( \p_{\ta}^2 +\cot \ta \p_\ta  \Big),\\
&& \nabla^2 \textbf{P} = \hat{r}\nabla^2 p + \hat{\ta}\nabla^2 q + O\left(\frac{1}{r^2}\right).
\end{eqnarray}
\end{subequations}
We define the iso-surface of the phagocyte interface as
\begin{equation}\label{ex-rho}
  u\Big( r=\rho(\ta,t),\ta,t \Big)=\frac{1}{2} .
\end{equation}

In Appendix A we calculate in details the projection operator term,
\begin{subequations}\label{Proj}
\begin{eqnarray}
&& \hat{P}\nabla u = \frac{1}{N}\Bigg\{ \left[ d^2 \sin^2 \ta u_r - d\sin\ta(r-d\cos\ta)\frac{u_\ta}{r} \right]\hat{r}+ \nonumber \\
&& \left[ -d\sin\ta(r-d\cos\ta)u_r + (r-d\cos\ta)^2  \frac{u_\ta}{r} \right]\hat{\ta} \Bigg\},\\
&& d=r_0 + R_0, \quad N = r^2-2rd\cos\ta + d^2.
\end{eqnarray}
\end{subequations}

 In order to balance the front dynamics with curvature we impose the following  scaling that describes slow dynamics
of a large-size cell \cite{AbuHamed2020},
\begin{equation}\label{scaling}
  \tilde{t}=\epsilon^2 t, \quad \rho(\ta,t)=\epsilon^{-1} R(\ta,t), \quad \eps\ll1 .
\end{equation}
The transition zone variable is defined as
\begin{equation}\label{zeta}
\z=r-\rho(\ta,t)=O(1).
\end{equation}
Also we define,
\begin{subequations}\label{field-z}
\begin{eqnarray}
 &&  R(\ta,t) = \tilde{R}(\ta, \tilde{t}) , \   u(r,\ta,t) = \tilde{u}(\z,\ta,\tilde{t}), \\
 &&  \textbf{P}(r,\ta,t)=\tilde{\textbf{P}}(\z,\ta,\tilde{t}).
\end{eqnarray}
\end{subequations}
The chain rule yields
\begin{equation}\label{chain1}
  \p_t = -\eps  \tilde{R}_{\tilde{t}}\p_\z + \eps^2 \p_{\tilde{t}}, \quad \p_r = \p_\z.
\end{equation}
Later on we drop the tildes.
 It holds that
 \begin{subequations}\label{chain2}
 \begin{eqnarray}
 && \frac{1}{r} = \frac{\epsilon}{R} - \frac{\epsilon^2 \z}{R^2 } +..., \\
 && \Psi_p(\textbf{r})\sim \exp\left( -\frac{G^2}{\tau D_p} \right),\\
 && \nabla \Psi_u(\textbf{r}) \sim \frac{-2\eps}{D_u} G \exp(-G^2 /D_u) \left[ G_R \hat{r} +\frac{G_\ta}{R} \hat{\ta} \right],\\
 && G(R,\ta)= \nonumber \\
 && \sqrt{R^2 -2RR_0(1+\lambda)\cos\ta + R_0^2 (1+\lambda)^2 } -\lambda R_0.
 \end{eqnarray}
\end{subequations}
In addition one can calculate,
\begin{subequations}
\begin{eqnarray}
&& \p_\ta = -\eps^{-1} R_\ta \p_\z + \p_\ta,\\
&& \p_\ta^2  = \eps^{-2} R_\ta^2 \p_{\z}^2 - \eps^{-1}( R_{\ta\ta}\p_\z +2R_\ta \p_{\z\ta}^2 ) + \p_{\ta}^2,\\
&& \frac{\p_\ta}{r} = -\frac{R_\ta}{R} \p_\z +O(\eps),\  \frac{\p_\ta}{r^2} = -\eps \frac{R_\ta}{R^2} \p_\z +O(\eps^2),\\
&& \nabla^2 u = \left( 1 + \frac{R_\ta^2}{R^2} \right) u_{\z\z} + \nonumber\\
&& \eps \Bigg[   \left( \frac{2}{R} - \frac{R_\ta}{R^2}\cot \ta - \frac{R_{\ta\ta}}{R^2} \right)u_\z \nonumber\\
&&  -\frac{2 R_\ta}{R^2} u_{\z\ta}   -\frac{2\z}{R^3} R_\ta^2   u_{\z\z}  \Bigg]+O(\eps^2). \label{LO}
\end{eqnarray}
\end{subequations}
We can approximate the  nonlocality in (\ref{Mod2}) as follows,
 \begin{equation}\label{nonloc}
   \int u \di^3 r \sim   \frac{2\pi\eps^{-3}}{3} \int_{0}^{\pi} R^3 (\ta,t) \sin\ta  \di \ta .
 \end{equation}
Consider the following scaling of the model parameters
\begin{eqnarray}\label{scal1}
 &&\alpha = \eps A, \quad   \frac{4\pi\mu}{3}\eps^{-3} = \eps M, \quad \sigma = \eps S, \nonumber\\
 &&  k=O(1), \quad \lambda=O(1) .
\end{eqnarray}
  Let us introduce the expansions
 \begin{equation}\label{expan}
  u = u_0 + \eps  u_1+..., \quad p = p_0 + \eps p_1 +...
\end{equation}
We define the auxiliary function, 
\begin{equation}\label{Lambda}
   \Lambda(\ta,t)= \left( 1+ \frac{R_\ta^2}{R^2}  \right)^{-1/2} ,
 \end{equation}
 and the function,
 \begin{eqnarray*}
   &&\Phi(\tau,D_u , D_p, \z  ) =\label{Phi(z)}\\
  &&\frac{1}{8}\sqrt{\frac{\tau}{2 D_u D_p}} \int_{-\infty}^{\infty} \ex^{-|s|/\sqrt{\tau D_p}} \cosh^{-2} \left( \frac{s - \z}{\sqrt{8D_u}} \right) \di s, \nonumber
 \end{eqnarray*}
 that are basic for our next analysis.

 We substitute the length, time \eqref{scaling}, and the parameter scaling \eqref{scal1}  into system \eqref{Mod}. We
write the system \eqref{Mod} using the transition zone variables \eqref{zeta}-\eqref{field-z}, and the chain rules \eqref{chain1}-\eqref{nonloc}.  We substitute the asymptotic expansions \eqref{expan} and finally we collect terms of the same order.

The equation at the leading order for $u$ is:
\begin{eqnarray}
&& D_u \Lambda^{-2} u_{0\z\z} = (1-u_0)(\frac{1}{2}-u_0)u_0,\\
&& u_0(\z\rightarrow -\infty)=1, \quad u_0(\z\rightarrow\infty)=0.
\end{eqnarray}
The Ginzburg-Landau theory yields,
\begin{equation}\label{}
 u_0 (\z) = \frac{1}{2} \left[ 1-\tanh\left(\frac{\Lambda\z}{\sqrt{8D_u}}\right) \right].
\end{equation}
\begin{figure}
  \centering
  \includegraphics[scale=0.3]{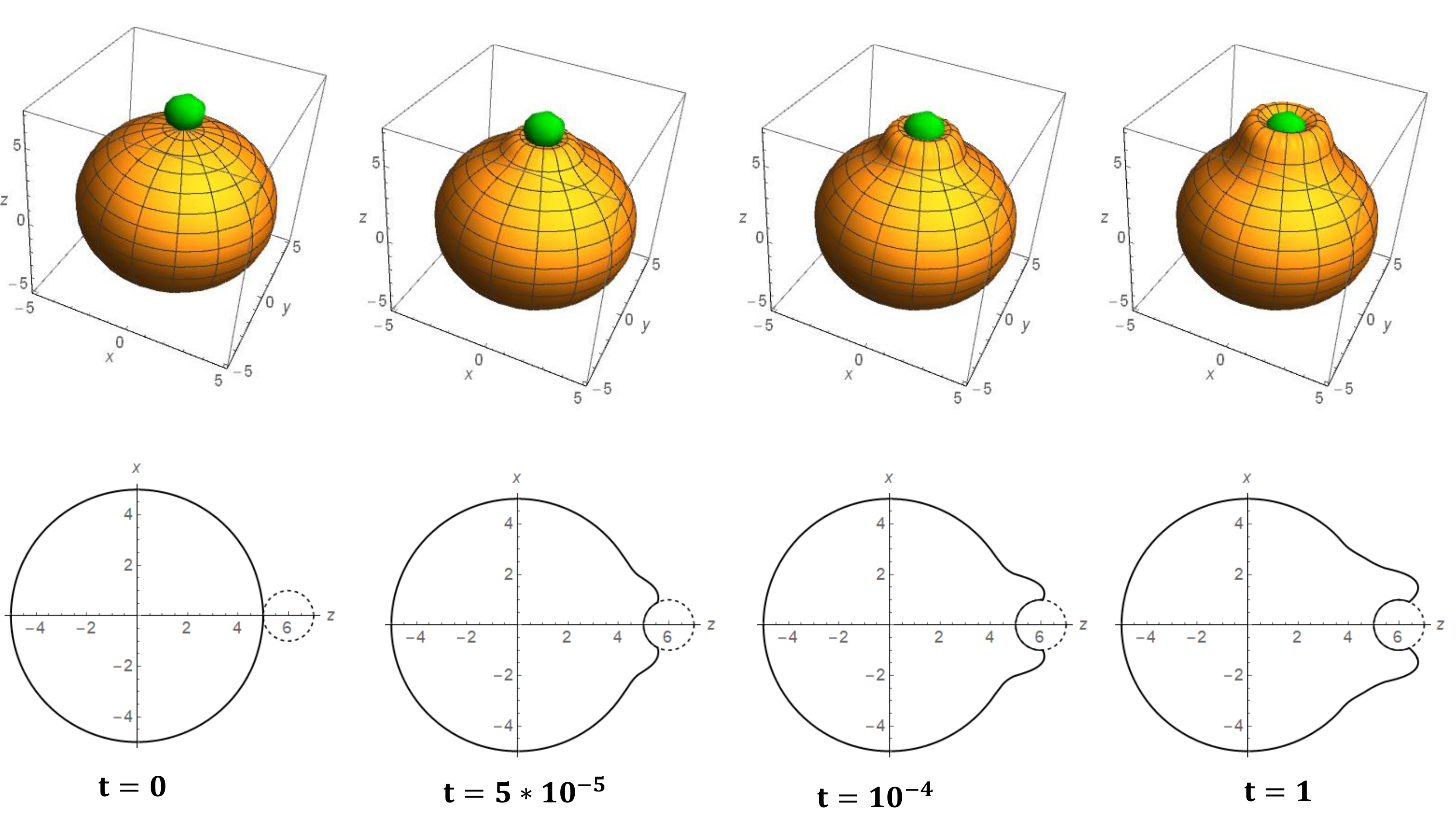}
  \caption{ Simulation of the evolutionary  equation of the phagocyte membrane \eqref{eveq} for time sequence mentioned in each of the engulfment process of the target particle (green sphere). Upper row is the spherical plot, lower row is the polar plot of the function $R(\ta,t)$.   The simulation manifests the pedestal-phase and then the cup phase but not the rounding phase.  We consider an initial sphere of radius $R_0=5$ , and spherical target particle with radius $r_0=\lambda R_0=1$, where $\lambda=0.2$. We use the following values of parameters $\beta=100, A=2, \tau=10,D_u=1,D_p=0.2,M=8,S=1,\nu=0.2, k=15$.     } \label{phag1}
\end{figure}
The equation at the next order $O(\eps)$ for $u_1$ is:
\begin{eqnarray}\label{}
&& L[u_1]=\textrm{RHS},\label{curef}\\
&&L =  D_u \Lambda^{-2} \p_\z^2  - \left(\frac{1}{2} -3 u_0 + 3u_0^2 \right)\hat{I},\nonumber\\
&& \textrm{RHS}=  - R_t u_{0\z}- D_u \Bigg[ \left( \frac{2}{R} - \frac{R_\ta}{R^2}\cot\ta - \frac{R_{\ta\ta}}{R^2} \right)u_{0\z} \nonumber\\
&&  -\frac{2 R_\ta}{R^2}  u_{0\z\ta} -\frac{2\z R_\ta^2}{R^3}   u_{0\z\z}  \Bigg]+\nonumber\\
&& (1-u_0)u_0 \left[  \tilde{V}(t) -  S(p_0^2 + q_0^2 )  \right]+  \nonumber\\
&&  A  \left( p_0 -\frac{R_\ta}{R} q_0    \right)u_{0\z}   - \nonumber\\
&& \frac{2k}{D_u} G  \exp(-G^2  /D_u) \left( G_R - \frac{G_\ta R_\ta}{R^2} \right)   u_{0\z} \nonumber
\end{eqnarray}
 where the volume variation have the form,
 \begin{equation*}
  \tilde{V} (t)= M \left[ \frac{1}{2} \int_{0}^{\pi} R^3 (\ta,t) \sin\ta  \di \ta - R_0^3 \right].
 \end{equation*}
The equation for the polarization field at the leading order:
\begin{eqnarray}\label{}
&& D_p \Lambda^{-2} p_{0\z\z} - \tau^{-1} p_0 =\beta \lambda_p u_{0\z},  \nonumber\\
&& \lambda_p = \exp\left(-\frac{G^2}{\tau D_p}\right)[ (1-\nu)gg_1 +\nu],\\
&& D_p \Lambda^{-2} q_{0\z\z} - \tau^{-1} q_0 = -\beta \lambda_q u_{0\z},\nonumber\\
&&\lambda_q = \exp\left(-\frac{G^2}{\tau D_p}\right) \left[(1-\nu) gg_2 + \nu\frac{R_\ta}{R} \right],
\end{eqnarray}
that could be solved via Fourier transform,
\begin{eqnarray}\label{}
&& p_0 (\z) =   \beta \lambda_p  \Lambda\Phi(\Lambda\z), \quad q_0 (\z) =  -\beta \lambda_q \Lambda\Phi(\Lambda\z)
\end{eqnarray}
 For the definition of $g,g_1,g_2$ see Appendix \ref{A}. We apply the solvability condition, which is the orthogonality of the  right-hand side (RHS) of equation \eqref{curef} to the solution $u_{0\z}$ of the homogenous equation $L[u]=0$ of \eqref{curef} i.e.,
\begin{equation}\label{}
\int_{-\infty}^{\infty}\textrm{RHS} (\z)\cdot u_{0\z} (\z) \di \z =0.
\end{equation}
We therefore obtain a closed evolution equation for the phagocyte interface $R(\ta,t)$:
\begin{eqnarray}\label{eveq}
&& a \Lambda R_t = -2a D_u \mathcal{H} - \tilde{V} + \Omega - E,
\end{eqnarray}
 where
 \begin{equation}\label{curv}
 \mathcal{H} =\frac{1}{2} \nabla\cdot \hat{n} = \frac{1}{2} \nabla\cdot \left( \frac{\nabla(r-R)}{|\nabla(r-R)|} \right)
 \end{equation}
 is the mean local curvature of the surface $r= R(\ta,t)$, see Appendix \ref{B}, and
 \begin{eqnarray}
 && a = \frac{1}{\sqrt{2 D_u}} \nonumber\\
&& \Omega(\ta,\ph,t) = 6\beta A \Omega_1 \Lambda^2 \left(\lambda_p + \frac{R_\ta}{R}\lambda_q  \right)+\nonumber\\
&& 6\beta^2 S \Omega_2 \Lambda^2  \left(\lambda_p^2 + \lambda_q^2  \right),\\
&& \Omega_1 (\tau ,D_u , D_p  ) = \int_{-\infty}^{\infty} \Phi(\xi) \bar{u}_{0\xi}^2(\xi) \di \xi, \\
&& \Omega_2 (\tau, D_u , D_p  ) = \nonumber \\
&&\int_{-\infty}^{\infty} \Phi^2 (\xi) (\bar{u}_0(\xi)-1) \bar{u}_0(\xi) \bar{u}_{0\xi}(\xi) \di \xi>0,
 \end{eqnarray}
 where
 \begin{equation}\label{}
\bar{u}_0 (\xi) = \frac{1}{2} \left[ 1-\tanh\left(\frac{\xi}{\sqrt{8D_u}}\right) \right].
 \end{equation}
 The adhesion effect is implemented by
 \begin{equation}\label{}
   E= \frac{2ak\Lambda}{D_u} G  \exp(-G^2  /D_u) \left( G_R - \frac{G_\ta R_\ta}{R^2} \right).
 \end{equation}
 Equation \eqref{eveq} is solved in conjunctions with initial and boundary conditions that guarantee a regular and
smooth solution,
  \begin{subequations}\label{bc}
  \begin{eqnarray}
  && R(t=0)=R_0,\\
  && R(\ta=0)=R_0, \quad R_\ta(\ta=0)=0, \\
  && R_\ta (\ta=\pi)=0 .
 \end{eqnarray}
 \end{subequations}
 Recall that following the definitions in \cite{Winkler+Aranson+Ziebert2019}, the interface of the cell is a diffuse
interface.
In order to prevent the penetration of the phagocyte interface inside the rigid bead, we add an additional constraint:
the motion of the phagocyte surface is stopped in the points where the distance between the interface and the bead
center is equal to $r_0$, i.e. $R(\ta,t)$ satisfy the inequality $X^2 +Y^2 +
(Z-(R_0 + r_0))^2 = r_0^2$; here $X^2+Y^2=R^2\sin^2\theta$, $Z=R\cos\theta$.

Equation \eqref{eveq} with initial and boundary conditions \eqref{bc}  are solved numerically via the function
\verb"NDSolve" of Wolfram Mathematica.
 In Fig. \ref{phag1} we present the simulation with initial phagocyte radius $R_0=5$, and bead radius $r_0=1$, hence $\lambda=0.2$.
Also, in Fig.
\ref{phag5} we carry out the same simulation for a twice bigger bead $\lambda=0.4$.

%


In some experiments phagocyte where expose to a bead that is much larger than the phagocyte thus the full engulfment is impossible in this case \cite{Herant+Heinrich+Dembo2006}. In Fig \ref{phag6} we present the steady state solution of the simulation of \eqref{eveq}, where  we choose $R_0=1$ and $\lambda=2$ i.e., the bead diameter is twice the phagocyte diameter. As expected we observe only the cup phase as in experiments \cite{Lee+Thompson2015}.

\section{Conclusion}
In this paper we attempted to introduce a simple model for phagocytosis using the phase field approach in the spirit of \cite{Winkler+Aranson+Ziebert2019}.
Our initial geometry is a phagocyte sphere tangent to a motionless rigid bead sphere. Phagocytosis is the remodeling or deformation of the phagocyte sphere around the bead Fig \ref{schematic}. We suggest a model that couples the order parameter $u$ with 3D polarization (orientation) vector field $\textbf{P}$ of the actin network of the phagocyte cytoskeleton \eqref{Mod}.

We derive a single scalar closed integro-differential equation governing the 3D phagocyte membrane dynamics \eqref{eveq} during bead engulfment, which includes the normal velocity $v_n=\Lambda R_t$ of the membrane curvature $\mathcal{H}$, volume relaxation rate $\tilde{V}$, a function $\Omega(t)$ determined by the molecular effects of the subcell level, and the adhesion effect $E$ of the motionless rigid spherical bead.

Our model is limited for several reasons.
We use the asymptotic assumption that $\eps\ll1$ to enables asymptotic analysis. Also, the
appearance of $\eps$
in the exponents in the definition of the static fields \eqref{Psi} in order to avoid boundary layer problem complications both in time and space.

This project is a primary attempt to utilize the phase field approach to model phagocytosis dynamics. There is a need for a future project with more computational depth, that consider the full computational model in \cite{Winkler+Aranson+Ziebert2019} without any asymptotic assumption, hopefully the full rounding phase, and the contact area power law that observed in \cite{Richards+Endres2014}, will recover.

%

\begin{appendices}
\section{}\label{A}
In this appendix we perform a detailed analysis for calculation of the projection operator $\hat{P}\nabla u$.
Denote
\begin{equation}\label{}
  d=R_0+r_0, \quad N= r^2 -2rd\cos\ta +d^2.
\end{equation}
The upward shifted sphere is given by:
\begin{equation}\label{}
  F= x^2 +y^2 +(z-d)^2 -r_0^2= r^2 -2dr\cos\ta + d^2 -r_0^2=0.
\end{equation}
Therefore,
\begin{equation}\label{}
  \nabla F = 2(r-d\cos\ta) \hat{r} + 2d\sin\ta \ \hat{\ta} .
\end{equation}
As a result we have the dyad:
\begin{eqnarray}\label{}
  && \hat{n}\hat{n}=\frac{\nabla F \nabla F}{|| \nabla F ||^2} = \frac{1}{N} \Big[ (r-d\cos\ta)^2 \hat{r}\hat{r} +  \\
  && d\sin\ta (r-d\cos\ta) (\hat{r}\hat{\ta}+\hat{\ta}\hat{r})+   d^2 \sin^2 \ta \ \hat{\ta}\hat{\ta} \Big]
   \end{eqnarray}
 Consequently we have the expression \eqref{Proj} for $\hat{P}\nabla u$.
 The asymptotic expansion at the leading order:
\begin{eqnarray}\label{}
  &&\hat{P}\nabla u \sim g(R,\ta) u_{0\z}\Big\{ \Big[ R_0^2 (1+\lambda)^2 \sin^2 \ta + \nonumber \\
  && R_0 (1+\lambda)\sin\ta (R-R_0 (1+\lambda)\cos\ta ) \frac{R_\ta}{R} \Big]\hat{r} - \nonumber\\
  && \Big[ R_0 (1+\lambda)\sin\ta (R-R_0 (1+\lambda)\cos\ta ) + \nonumber\\
  &&  (R-R_0 (1+\lambda)\cos\ta )^2 \frac{R_\ta}{R} \Big]\hat{\ta}\Big\}= \nonumber\\
  &&  g(R,\ta) u_{0\z}\{ g_1(R,\ta)\hat{r} - g_2 (R,\ta) \hat{\ta} \}
   \end{eqnarray}
where
\begin{subequations}\label{g}
\begin{eqnarray}
 && g (R,\ta)= \nonumber \\
 && 1/[R^2 - 2RR_0(1+\lambda)\cos\ta + R_0^2 (1+\lambda)^2 ],\\
 && g_1(R,\ta) = R_0^2 (1+\lambda)^2 \sin^2 \ta + \nonumber\\
 &&  R_0 (1+\lambda)\sin\ta (R-R_0 (1+\lambda)\cos\ta ) \frac{R_\ta}{R} ,\\
 && g_2(R,\ta) = R_0 (1+\lambda)\sin\ta (R-R_0 (1+\lambda)\cos\ta ) +\nonumber\\
 &&  (R-R_0 (1+\lambda)\cos\ta )^2 \frac{R_\ta}{R}.
\end{eqnarray}
\end{subequations}

\section{}\label{B}
Here we give an explicit expression for the mean curvature of a surface given in spherical coordinate description
$r=R(\ta,t)$, assuming a cylindrical symmetry ,see Fig. \ref{schematic}. Following the definitions \eqref{coor1}, and
\eqref{Lambda}
one can calculate,
\begin{eqnarray}\label{}
  &&  \nabla \cdot \hat{n} = \Lambda \Bigg[ \frac{2}{R} - \frac{R_{\ta\ta}}{R^2} - \frac{R_\ta \cot\ta}{R^2}  \Bigg] +\nonumber\\
  && \Lambda^3 \Bigg[ \frac{R_\ta^2}{R^3} + \frac{R_\ta^2 R_{\ta\ta}}{R^4}    \Bigg]
 \end{eqnarray}

\end{appendices}

\begin{figure}
  \centering
  \includegraphics[scale=0.3]{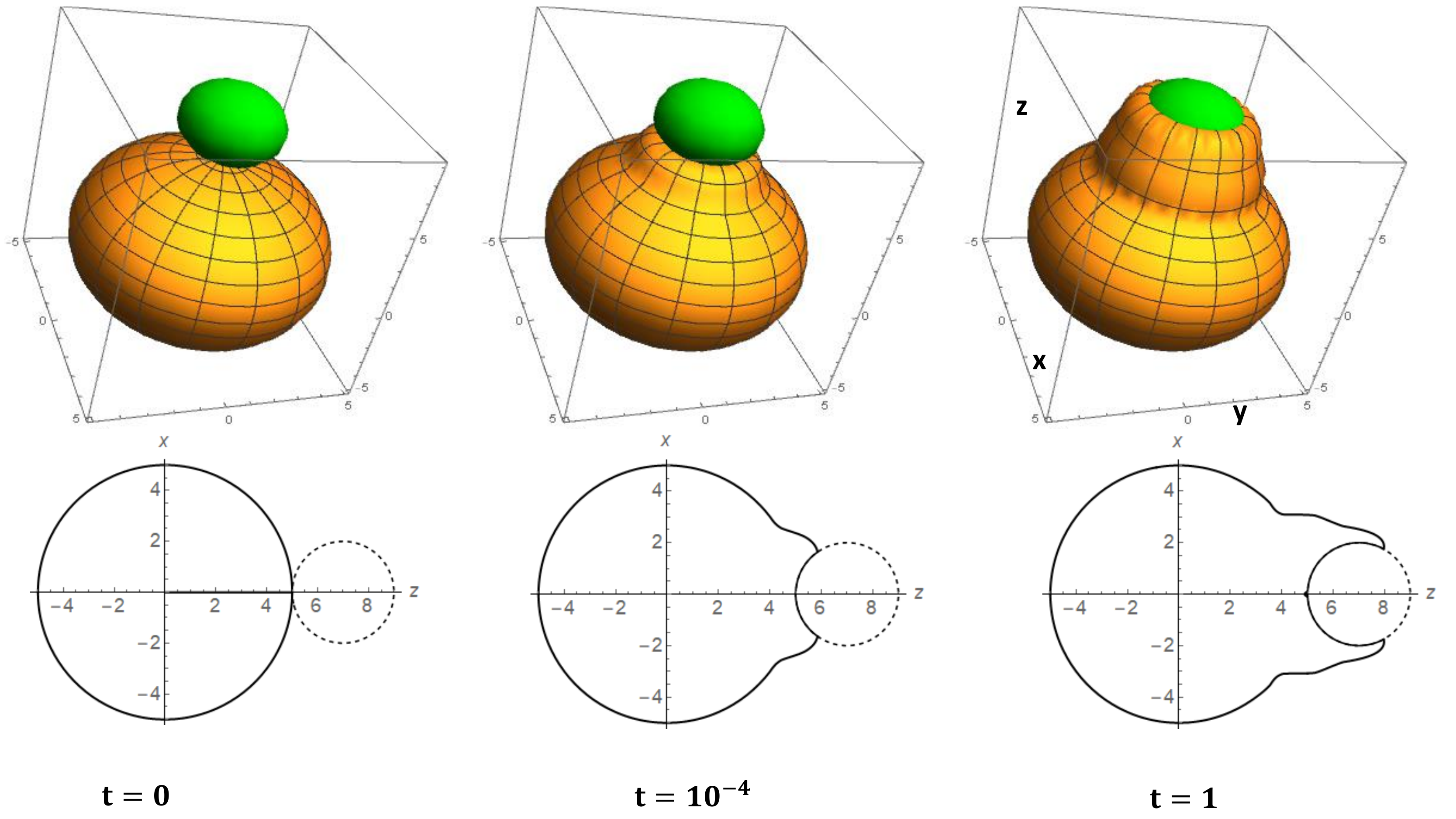}
  \caption{ Simulation of the evolutionary  equation of the phagocyte membrane \eqref{eveq} for time sequence mentioned in each of the engulfment process of a small target particle (green sphere). We consider the same value of parameters as in Fig. \ref{phag1}, except $\lambda=0.4,$ .     } \label{phag5}
\end{figure}
\begin{figure}
  \centering
  \includegraphics[scale=0.25]{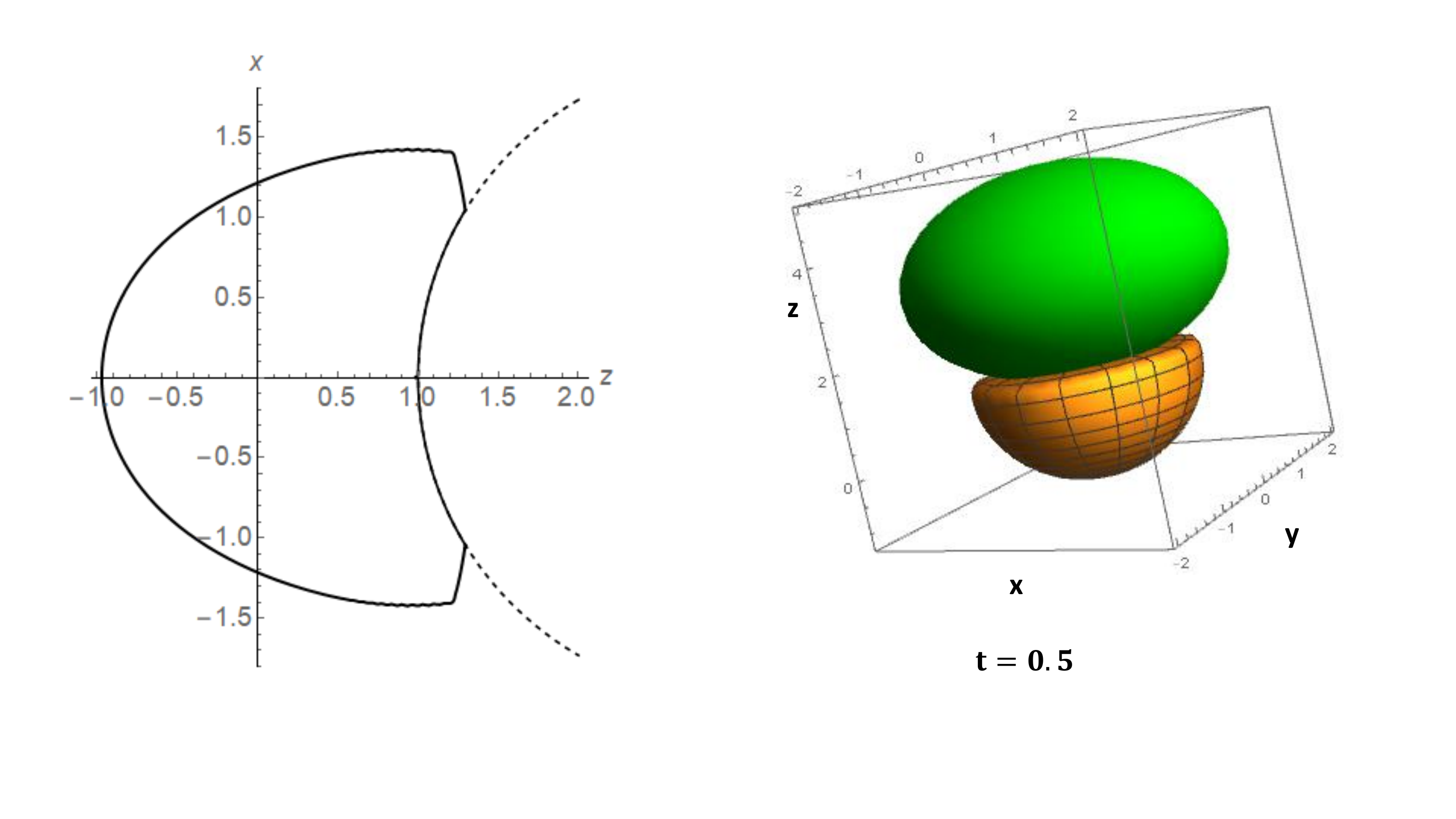}
  \caption{ Simulation of the evolutionary  equation of the phagocyte membrane \eqref{eveq} for time $t=0.5$ . Left is
the polar plot and right is the spherical plot of the function $R(\ta,t=0.5)$.  We consider the same value of
parameters as in Fig. \ref{phag1}, except $R_0=1, \beta=10, \lambda=2$, and $M=20$.
   } \label{phag6}

\end{figure}

\clearpage
\bibliography{Phag-refs}{}

\begin{thebibliography}{16}%
\makeatletter
\providecommand \@ifxundefined [1]{%
 \@ifx{#1\undefined}
}%
\providecommand \@ifnum [1]{%
 \ifnum #1\expandafter \@firstoftwo
 \else \expandafter \@secondoftwo
 \fi
}%
\providecommand \@ifx [1]{%
 \ifx #1\expandafter \@firstoftwo
 \else \expandafter \@secondoftwo
 \fi
}%
\providecommand \natexlab [1]{#1}%
\providecommand \enquote  [1]{``#1''}%
\providecommand \bibnamefont  [1]{#1}%
\providecommand \bibfnamefont [1]{#1}%
\providecommand \citenamefont [1]{#1}%
\providecommand \href@noop [0]{\@secondoftwo}%
\providecommand \href [0]{\begingroup \@sanitize@url \@href}%
\providecommand \@href[1]{\@@startlink{#1}\@@href}%
\providecommand \@@href[1]{\endgroup#1\@@endlink}%
\providecommand \@sanitize@url [0]{\catcode `\\12\catcode `\$12\catcode
  `\&12\catcode `\#12\catcode `\^12\catcode `\_12\catcode `\%12\relax}%
\providecommand \@@startlink[1]{}%
\providecommand \@@endlink[0]{}%
\providecommand \url  [0]{\begingroup\@sanitize@url \@url }%
\providecommand \@url [1]{\endgroup\@href {#1}{\urlprefix }}%
\providecommand \urlprefix  [0]{URL }%
\providecommand \Eprint [0]{\href }%
\providecommand \doibase [0]{http://dx.doi.org/}%
\providecommand \selectlanguage [0]{\@gobble}%
\providecommand \bibinfo  [0]{\@secondoftwo}%
\providecommand \bibfield  [0]{\@secondoftwo}%
\providecommand \translation [1]{[#1]}%
\providecommand \BibitemOpen [0]{}%
\providecommand \bibitemStop [0]{}%
\providecommand \bibitemNoStop [0]{.\EOS\space}%
\providecommand \EOS [0]{\spacefactor3000\relax}%
\providecommand \BibitemShut  [1]{\csname bibitem#1\endcsname}%
\let\auto@bib@innerbib\@empty
\bibitem [{\citenamefont {Winkler}\ \emph {et~al.}(2019)\citenamefont
  {Winkler}, \citenamefont {Aranson},\ and\ \citenamefont
  {Ziebert}}]{Winkler+Aranson+Ziebert2019}%
  \BibitemOpen
  \bibfield  {author} {\bibinfo {author} {\bibfnamefont {B.}~\bibnamefont
  {Winkler}}, \bibinfo {author} {\bibfnamefont {I.~S.}\ \bibnamefont
  {Aranson}}, \ and\ \bibinfo {author} {\bibfnamefont {F.}~\bibnamefont
  {Ziebert}},\ }\bibfield  {title} {\enquote {\bibinfo {title} {Confinement and
  substrate topography control cell migration in a 3d computational model},}\
  }\href@noop {} {\bibfield  {journal} {\bibinfo  {journal} {Communications
  Physics}\ }\textbf {\bibinfo {volume} {2:82}} (\bibinfo {year}
  {2019})}\BibitemShut {NoStop}%
\bibitem [{\citenamefont {Flannagan}\ \emph {et~al.}(2012)\citenamefont
  {Flannagan}, \citenamefont {Jaumouille},\ and\ \citenamefont
  {Grinstein}}]{Flannagan+Jaumouille+Grinstein2012}%
  \BibitemOpen
  \bibfield  {author} {\bibinfo {author} {\bibfnamefont {R.~S.}\ \bibnamefont
  {Flannagan}}, \bibinfo {author} {\bibfnamefont {V.}~\bibnamefont
  {Jaumouille}}, \ and\ \bibinfo {author} {\bibfnamefont {S.}~\bibnamefont
  {Grinstein}},\ }\bibfield  {title} {\enquote {\bibinfo {title} {The cell
  biology of phagocytosis},}\ }\href@noop {} {\bibfield  {journal} {\bibinfo
  {journal} {The Annual Review of Pathology: Mechanisms of Disease}\ }\textbf
  {\bibinfo {volume} {7}},\ \bibinfo {pages} {61–98} (\bibinfo {year}
  {2012})}\BibitemShut {NoStop}%
\bibitem [{\citenamefont {Poon}\ \emph {et~al.}(2010)\citenamefont {Poon},
  \citenamefont {Hulett},\ and\ \citenamefont
  {Parish}}]{Poon+Hulett+Parish2010}%
  \BibitemOpen
  \bibfield  {author} {\bibinfo {author} {\bibfnamefont {I.}~\bibnamefont
  {Poon}}, \bibinfo {author} {\bibfnamefont {MD.}\ \bibnamefont {Hulett}}, \
  and\ \bibinfo {author} {\bibfnamefont {CR}~\bibnamefont {Parish}},\
  }\bibfield  {title} {\enquote {\bibinfo {title} {Molecular mechanisms of late
  apoptotic/necrotic cell clearance},}\ }\href@noop {} {\bibfield  {journal}
  {\bibinfo  {journal} {Cell Death and Differentiation}\ }\textbf {\bibinfo
  {volume} {17}},\ \bibinfo {pages} {381–397} (\bibinfo {year}
  {2010})}\BibitemShut {NoStop}%
\bibitem [{\citenamefont {Querol}\ and\ \citenamefont
  {Rosales}(2020)}]{Querol+Rosales2020}%
  \BibitemOpen
  \bibfield  {author} {\bibinfo {author} {\bibfnamefont {E.~U.}\ \bibnamefont
  {Querol}}\ and\ \bibinfo {author} {\bibfnamefont {C.}~\bibnamefont
  {Rosales}},\ }\bibfield  {title} {\enquote {\bibinfo {title} {Phagocytosis:
  Our current understanding of a universal biological process},}\ }\href@noop
  {} {\bibfield  {journal} {\bibinfo  {journal} {Front. Immunol.}\ }\textbf
  {\bibinfo {volume} {11}} (\bibinfo {year} {2020})}\BibitemShut {NoStop}%
\bibitem [{\citenamefont {Rosales}\ and\ \citenamefont
  {Querol}(2017)}]{Rosales+Querol2017}%
  \BibitemOpen
  \bibfield  {author} {\bibinfo {author} {\bibfnamefont {C.}~\bibnamefont
  {Rosales}}\ and\ \bibinfo {author} {\bibfnamefont {E.~U.}\ \bibnamefont
  {Querol}},\ }\bibfield  {title} {\enquote {\bibinfo {title} {Phagocytosis: A
  fundamental process in immunity},}\ }\href@noop {} {\bibfield  {journal}
  {\bibinfo  {journal} {BioMed Research International}\ }\textbf {\bibinfo
  {volume} {2017}},\ \bibinfo {pages} {9042851} (\bibinfo {year}
  {2017})}\BibitemShut {NoStop}%
\bibitem [{\citenamefont {Herant}\ \emph {et~al.}(2006)\citenamefont {Herant},
  \citenamefont {Heinrich},\ and\ \citenamefont
  {Dembo}}]{Herant+Heinrich+Dembo2006}%
  \BibitemOpen
  \bibfield  {author} {\bibinfo {author} {\bibfnamefont {M.}~\bibnamefont
  {Herant}}, \bibinfo {author} {\bibfnamefont {V.}~\bibnamefont {Heinrich}}, \
  and\ \bibinfo {author} {\bibfnamefont {M.}~\bibnamefont {Dembo}},\ }\bibfield
   {title} {\enquote {\bibinfo {title} {Mechanics of neutrophil phagocytosis:
  experiments and quantitative models},}\ }\href@noop {} {\bibfield  {journal}
  {\bibinfo  {journal} {Journal of Cell Science}\ }\textbf {\bibinfo {volume}
  {119}},\ \bibinfo {pages} {1903--1913} (\bibinfo {year} {2006})}\BibitemShut
  {NoStop}%
\bibitem [{\citenamefont {Richards}\ and\ \citenamefont
  {Endres}(2014)}]{Richards+Endres2014}%
  \BibitemOpen
  \bibfield  {author} {\bibinfo {author} {\bibfnamefont {D.~M.}\ \bibnamefont
  {Richards}}\ and\ \bibinfo {author} {\bibfnamefont {R.~G.}\ \bibnamefont
  {Endres}},\ }\bibfield  {title} {\enquote {\bibinfo {title} {The mechanism of
  phagocytosis: Two stages of engulfment},}\ }\href@noop {} {\bibfield
  {journal} {\bibinfo  {journal} {Biophysical Journal}\ }\textbf {\bibinfo
  {volume} {107}} (\bibinfo {year} {2014})}\BibitemShut {NoStop}%
\bibitem [{\citenamefont {Tollis}\ \emph {et~al.}(2010)\citenamefont {Tollis},
  \citenamefont {Dart}, \citenamefont {Tzircotis},\ and\ \citenamefont
  {Endres}}]{Tollis+Dart+Tzircotis+Endres2010}%
  \BibitemOpen
  \bibfield  {author} {\bibinfo {author} {\bibfnamefont {S.}~\bibnamefont
  {Tollis}}, \bibinfo {author} {\bibfnamefont {A.~E.}\ \bibnamefont {Dart}},
  \bibinfo {author} {\bibfnamefont {G.}~\bibnamefont {Tzircotis}}, \ and\
  \bibinfo {author} {\bibfnamefont {R.~G.}\ \bibnamefont {Endres}},\ }\bibfield
   {title} {\enquote {\bibinfo {title} {The zipper mechanism in phagocytosis:
  energetic requirements and variability in phagocytic cup shape},}\
  }\href@noop {} {\bibfield  {journal} {\bibinfo  {journal} {BMC Systems
  Biology}\ }\textbf {\bibinfo {volume} {4}} (\bibinfo {year}
  {2010})}\BibitemShut {NoStop}%
\bibitem [{\citenamefont {van Zon}\ \emph {et~al.}(2009)\citenamefont {van
  Zon}, \citenamefont {Tzircotis}, \citenamefont {Caron},\ and\ \citenamefont
  {Howard}}]{Zon+Tzircotis+Caron+Howard2009}%
  \BibitemOpen
  \bibfield  {author} {\bibinfo {author} {\bibfnamefont {J.~S.}\ \bibnamefont
  {van Zon}}, \bibinfo {author} {\bibfnamefont {G.}~\bibnamefont {Tzircotis}},
  \bibinfo {author} {\bibfnamefont {E.}~\bibnamefont {Caron}}, \ and\ \bibinfo
  {author} {\bibfnamefont {M.}~\bibnamefont {Howard}},\ }\bibfield  {title}
  {\enquote {\bibinfo {title} {A mechanical bottleneck explains the variation
  in cup growth during \uppercase{F}c$\gamma$\uppercase{R} phagocytosis},}\
  }\href@noop {} {\bibfield  {journal} {\bibinfo  {journal} {Molecular Systems
  Biology}\ }\textbf {\bibinfo {volume} {5}} (\bibinfo {year}
  {2009})}\BibitemShut {NoStop}%
\bibitem [{\citenamefont {Herant}\ \emph {et~al.}(2011)\citenamefont {Herant},
  \citenamefont {Lee}, \citenamefont {Dembo},\ and\ \citenamefont
  {Heinrich}}]{Herant+Lee+Dembo+Heinrich2011}%
  \BibitemOpen
  \bibfield  {author} {\bibinfo {author} {\bibfnamefont {M.}~\bibnamefont
  {Herant}}, \bibinfo {author} {\bibfnamefont {C.~Y.}\ \bibnamefont {Lee}},
  \bibinfo {author} {\bibfnamefont {M.}~\bibnamefont {Dembo}}, \ and\ \bibinfo
  {author} {\bibfnamefont {V.}~\bibnamefont {Heinrich}},\ }\bibfield  {title}
  {\enquote {\bibinfo {title} {Protrusive push versus enveloping embrace:
  Computational model of phagocytosis predicts key regulatory role of
  cytoskeletal membrane anchors.}}\ }\href@noop {} {\bibfield  {journal}
  {\bibinfo  {journal} {PLoS Comput Biol}\ }\textbf {\bibinfo {volume} {7(1)}}
  (\bibinfo {year} {2011})}\BibitemShut {NoStop}%
\bibitem [{\citenamefont {Richards}\ and\ \citenamefont
  {Endres}(2016)}]{Richards+Endres2016}%
  \BibitemOpen
  \bibfield  {author} {\bibinfo {author} {\bibfnamefont {D.~M.}\ \bibnamefont
  {Richards}}\ and\ \bibinfo {author} {\bibfnamefont {R.~G.}\ \bibnamefont
  {Endres}},\ }\bibfield  {title} {\enquote {\bibinfo {title} {Target shape
  dependence in a simple model of receptor-mediated endocytosis and
  phagocytosis},}\ }\href@noop {} {\bibfield  {journal} {\bibinfo  {journal}
  {PNAS}\ }\textbf {\bibinfo {volume} {113}},\ \bibinfo {pages} {6113–6118}
  (\bibinfo {year} {2016})}\BibitemShut {NoStop}%
\bibitem [{\citenamefont {Jaumouille}\ and\ \citenamefont
  {Waterman}(2020)}]{Jaumouille+Waterman2020}%
  \BibitemOpen
  \bibfield  {author} {\bibinfo {author} {\bibfnamefont {V.}~\bibnamefont
  {Jaumouille}}\ and\ \bibinfo {author} {\bibfnamefont {C.~M.}\ \bibnamefont
  {Waterman}},\ }\bibfield  {title} {\enquote {\bibinfo {title} {Physical
  constraints and forces involved in phagocytosis},}\ }\href@noop {} {\bibfield
   {journal} {\bibinfo  {journal} {Front. Immunol.}\ }\textbf {\bibinfo
  {volume} {11}} (\bibinfo {year} {2020})}\BibitemShut {NoStop}%
\bibitem [{\citenamefont {May}\ and\ \citenamefont
  {Machesky}(2001)}]{May+Machesky2001}%
  \BibitemOpen
  \bibfield  {author} {\bibinfo {author} {\bibfnamefont {R.~C.}\ \bibnamefont
  {May}}\ and\ \bibinfo {author} {\bibfnamefont {L.~M.}\ \bibnamefont
  {Machesky}},\ }\bibfield  {title} {\enquote {\bibinfo {title} {Phagocytosis
  and the actin cytoskeleton},}\ }\href@noop {} {\bibfield  {journal} {\bibinfo
   {journal} {Journal of Cell Science}\ }\textbf {\bibinfo {volume} {114}},\
  \bibinfo {pages} {1061--1077} (\bibinfo {year} {2001})}\BibitemShut {NoStop}%
\bibitem [{\citenamefont {Hamed}\ and\ \citenamefont
  {Nepomnyashchy}(2021)}]{AbuHamed2021}%
  \BibitemOpen
  \bibfield  {author} {\bibinfo {author} {\bibfnamefont {M.~Abu}\ \bibnamefont
  {Hamed}}\ and\ \bibinfo {author} {\bibfnamefont {A.A.}\ \bibnamefont
  {Nepomnyashchy}},\ }\bibfield  {title} {\enquote {\bibinfo {title}
  {Three--dimensional phase field model for actin--based cell membrane
  dynamics},}\ }\href@noop {} {\bibfield  {journal} {\bibinfo  {journal} {to be
  puplished in: Mathematical Modelling of Natural Phenomena}\ } (\bibinfo
  {year} {2021})}\BibitemShut {NoStop}%
\bibitem [{\citenamefont {Hamed}\ and\ \citenamefont
  {Nepomnyashchy}(2020)}]{AbuHamed2020}%
  \BibitemOpen
  \bibfield  {author} {\bibinfo {author} {\bibfnamefont {M.~Abu}\ \bibnamefont
  {Hamed}}\ and\ \bibinfo {author} {\bibfnamefont {A.A.}\ \bibnamefont
  {Nepomnyashchy}},\ }\bibfield  {title} {\enquote {\bibinfo {title} {A simple
  model of keratocyte membrane dynamics: The case of motionless living cell},}\
  }\href@noop {} {\bibfield  {journal} {\bibinfo  {journal} {Physica D}\
  }\textbf {\bibinfo {volume} {408}} (\bibinfo {year} {2020})}\BibitemShut
  {NoStop}%
\bibitem [{\citenamefont {Lee}\ \emph {et~al.}(2015)\citenamefont {Lee},
  \citenamefont {Thompson}, \citenamefont {Hastey}, \citenamefont {Hodge},
  \citenamefont {Lunetta}, \citenamefont {Pappagianis},\ and\ \citenamefont
  {Heinrich}}]{Lee+Thompson2015}%
  \BibitemOpen
  \bibfield  {author} {\bibinfo {author} {\bibfnamefont {C.~Y.}\ \bibnamefont
  {Lee}}, \bibinfo {author} {\bibfnamefont {G.~R.}\ \bibnamefont {Thompson}},
  \bibinfo {author} {\bibfnamefont {C.~J.}\ \bibnamefont {Hastey}}, \bibinfo
  {author} {\bibfnamefont {G.~C.}\ \bibnamefont {Hodge}}, \bibinfo {author}
  {\bibfnamefont {J.~M.}\ \bibnamefont {Lunetta}}, \bibinfo {author}
  {\bibfnamefont {D.}~\bibnamefont {Pappagianis}}, \ and\ \bibinfo {author}
  {\bibfnamefont {V.}~\bibnamefont {Heinrich}},\ }\bibfield  {title} {\enquote
  {\bibinfo {title} {Coccidioides endospores and spherules draw strong
  chemotactic, adhesive, and phagocytic responses by individual human
  neutrophils},}\ }\href@noop {} {\bibfield  {journal} {\bibinfo  {journal}
  {PLoS ONE}\ }\textbf {\bibinfo {volume} {10(6)}} (\bibinfo {year}
  {2015})}\BibitemShut {NoStop}%
\end{thebibliography}%

\end{document}